\documentclass{jfm}
\usepackage{graphicx}
\usepackage{epstopdf, epsfig}
\usepackage{color}

\shorttitle{Wetting states of two-dimensional drops under gravity}
\shortauthor{C. Lv}

\title{\bf Wetting states of two-dimensional drops under gravity}
\author
 {
 Cunjing Lv\aff{1}
  \corresp{\email{lu@nmf.tu-darmstadt.de}}
  }

\affiliation
{
\aff{1}
Institute for Nano- and Microfluidics, Center of Smart Interfaces, Technische Universit{\"a}t Darmstadt, Alarich-Weiss-Stra{\ss}e 10, 64287 Darmstadt, Germany
}

\begin{document}

\maketitle
% --
\begin{abstract}
An analytical model is proposed for the Young-Laplace equation of two-dimensional (2D) drops under gravity. Inspired by the pioneering work of \citet{Landau1987}, we derive analytical expressions of the profile of drops on flat surfaces, for arbitrary contact angles and drop volume. We then extend our theory for drops on inclined surfaces and reveal that the contact line plays a key role on the wetting state of the drops: (1) when the contact line is completely pinning, the advancing and receding contact angles and the shape of the drop can be uniquely determined by the predefined droplet volume, sliding angle and contact area, which does not rely on the Young contact angle; (2) when the drop has a movable contact line, it would achieve a wetting state with a minimum free energy resulting from the competition between the surface tension and gravity. Our theory is in excellent agreement with numerical results.
\end{abstract}
% --
% --------------------------------------------------------------------------------------------------
\section{Introduction}\label{Sec_1}
% --
When a drop is deposited on a surface, it adopts a specific shape which is governed by the Young-Laplace equation \citep{Young1805, Laplace1805}. Obtaining the solution of the Young-Laplace equation is fundamentally important for understanding the underlying physics of wetting, such as the capillary force, adhesion and friction at the solid-liquid interface, wetting transition, morphology of the liquid, etc. Influences such as the gravity and the roughness of the surface are of practical importance in wetting \citep{deGennes1985, Bonn2009, Lohse2015} and need to be taken into consideration. When gravity is considered, the exact (non-trivial) solutions of the Young-Laplace equation have only been found in the cases of: (1) a fluid in a semi-infinite domain bounded by a vertical plane wall; (2) or for a fluid between two vertical parallel walls. These results were both given by \citet{Landau1987} and they are solutions for wetting in two-dimensional (2D) space. Previously, researchers have resorted to approximate solutions to quantify the related questions, such as the shape of drops on flat surfaces \citep{Frenkel1948, Finn1986, Myshkis1987, Srinivasan2011}, pendant drops \citep{Michael1976, Chesters1977}, the balance between the surface tension and gravity for drops lying on inclined surfaces \citep{Frenkel1948, Furmidge1962, Olsen1962, Kim2002, Benilov2015}, meniscus/drop-on-fiber systems \citep{Clanet2002, deGennes2004}, the capillary rise in a wedge/tube \citep{Siegel1980, Wong1992, Fowkes1998, Norbury2004, Anderson2006}, etc. However, their utility has a limited scope because usually the contact angle $\theta$ or the effect of gravity (which is characterized by the Bond number) was assumed to be very small ($\mathrm{Bo} = \rho g l^2 / \sigma \ll 1$, denoting $\rho$, $g$, $l$ and $\sigma$ the density of the liquid, the gravitational acceleration, the size of the drop and the liquid-vapor surface tension). 

% --
When a drop is lying on an inclined surface in the presence of roughness, the question is more complicated. The only known exact relationship is for a 2D case \citep{Frenkel1948},
% ------------------
\begin{eqnarray}
\rho g V \sin{\alpha} = \sigma \left( \cos{\theta_{\rm R}} - \cos{\theta_{\rm A}} \right),
\label{Eq_1_1}
\end{eqnarray}
% ------------------
% --
\noindent
in which $\rho$ is the areal density of the drop with a cross-section $V$, $\theta_{\rm R}$ and $\theta_{\rm A}$ are the receding and advancing contact angles. $\alpha$ is the slope of the surface, when it reaches a critical value (sliding angle) the drop begins to slide down the surface. Eq. (\ref{Eq_1_1}) is simply built based on a force balance of different components of the surface tensions and gravity along the inclined surface. In section \ref{Sec_3}, we will verify that Eq. (\ref{Eq_1_1}) is essentially a boundary condition of the Young-Laplace equation. For a three-dimensional (3D) case, Eq. (\ref{Eq_1_1}) is modified to $\rho g V \sin \alpha = k w \sigma (\cos \theta_{\rm R} - \cos \theta_{\rm A})$, in which $w$ is the width of the solid-liquid contact area and $k$ is a numerical constant that depends on the shape of the drop \citep{Extrand1995}. Unfortunately, for given values of $\alpha$ and $V$, we cannot distinguish $\theta_{\rm R}$ and $\theta_{\rm A}$ from Eq. (\ref{Eq_1_1}) alone. Moreover, we cannot predict the sliding angle via Eq. (\ref{Eq_1_1}) with certain values of $V$ and $\theta$.

% --
So far, little information has been obtained about the exact solution of the Young-Laplace equation for drops under gravity. In the present study, we restrict our analysis to the 2D problem of drops, which is a natural extension of the seminal works on 2D wetting \citep{Frenkel1948, Olsen1962, Landau1987} and this simplification is easier to tackle than the 3D problem. In fact, 2D results have considerable practical applications to industrial problems, such as the dip-coating and printing processes, deposition and solidification of molten materials, anisotropic wettability on striped surfaces for fluidic control and transport \citep{Schiaffino1997, Gau1999, Xia2012, Reyssat2015}. Recently, interest in 2D geometry increases and some results suggest that the physics are almost indistinguishable between the 2D and 3D cases such as in liquid spreading, wettability of drops on soft solids, motion of long bubbles in channels \citep{Savva2010, Lubbers2014, Fabre2016}. Here, we deduce exact solutions of the Young-Laplace equation for 2D drops lying on both flat and inclined surfaces. We not only exactly determine all related quantities ($V$, $\alpha$, $\theta_{\rm R}$, $\theta_{\rm A}$, contact region, etc.) without any assumption or approximation, but also reveal the dependencies among them.
% --------------------------------------------------------------------------------------------------
%\clearpage
\section{General solution of the shape of drops lying on a horizontal surface}\label{Sec_2}
\begin{figure}
  \centerline{\includegraphics[width=8.5cm]{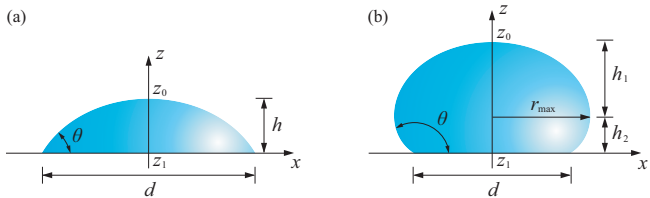}}
  \caption{Wetting states of 2D drops on horizontal surfaces under gravity. These profiles are obtained using Eq. (\ref{Eq_2_3}) and (\ref{Eq_2_4}): ({\it a}) $ \theta = 60^{\circ} $, $ d = 2a $; ({\it b}) $ \theta = 150^{\circ} $, $ d = 2a $ (the origin of the coordinate system is not at the center of the solid-liquid contact area, see figure \ref{fig:fig6}).}
\label{fig:fig1}
\end{figure}

% --
As shown in  figure \ref{fig:fig1}, we demonstrate the exact profiles of two drops lying on horizontal surfaces under gravity in 2D space. Practically, these shapes correspond to cross-sections of liquid on striped surfaces \citep{Xia2012}. The shape of the drop is governed by the Young-Laplace equation $\kappa \sigma = \Delta p$, where $\kappa$ and $\Delta p$ are the curvature and pressure difference between the liquid and vapour phases at any point of the meniscus. In figure \ref{fig:fig1}, the Young-Laplace equation can be expressed as,
% ------------------
\begin{eqnarray}
\frac{z^{\prime \prime}}{\left[ 1+ \left( z^{\prime} \right)^2 \right]^{3/2}} \cdot \sigma = \Delta p_0 + \rho g z,
\label{Eq_2_1}
\end{eqnarray}
% ------------------
% --
\noindent
in which $\Delta p_0$ is a constant. Previously,  researchers employed various approximate methods to solve Eq. (\ref{Eq_2_1}). The term $z^{\prime}$ was usually ignored (i.e., let $\kappa \approx z^{\prime \prime}$) and this view obtained a great success in the field of lubrication \citep{Bonn2009}, but the solution is limited to small contact angles. For high contact angles, researchers employed perturbation solutions \citep{Extrand2010, Srinivasan2011} and could also get good results. Even though, there is still a lack of comprehensive understanding, and a general solution which can be applied to any $\theta$ remains unaddressed.
% ************
% ************

% --
The exact solutions of the Young-Laplace equation under gravity obtained by \citet[pp. 242-243]{Landau1987} are only applicable to the profile of menisci bounded by one or two planes. To our best knowledge, it is the first time we deduce the exact solution of Eq. (\ref{Eq_2_1}) for 2D drops under gravity (see Appendix \ref{appA_1} and \ref{appA_2}). We obtain,
% ------------------
\begin{eqnarray}
V = 2a^2 \left[ \sqrt{A - \cos{\theta}} \int_0^{\theta} \frac{\cos{\xi}}{\sqrt{A - \cos{\xi}}} {\rm d}{\xi}  - \sin{\theta} \right],
\label{Eq_2_2}
\end{eqnarray}

\begin{eqnarray}
x = \pm \frac{\sqrt{2}a}{2} \int_0^{\eta} \frac{\cos{\xi}}{\sqrt{A - \cos{\xi}}} {\rm d}{\xi}, \quad \eta \in [0,\theta],
\label{Eq_2_3}
\end{eqnarray}

\begin{eqnarray}
z = - \sqrt{2} a \sqrt{A-\cos{\eta}}, \quad \eta \in [0,\theta],
\label{Eq_2_4}
\end{eqnarray}
% ------------------
% --
\noindent
where $\theta$ is defined using $\cos {\theta} = (\sigma_{\rm SV} - \sigma_{\rm SL}) / \sigma$ \citep{Young1805}, denoting $\sigma_{\rm SV}$ and $\sigma_{\rm SL}$ the solid-vapor and solid-liquid interfacial tensions. $a = (\sigma / \rho g)^{1/2}$ is the capillary length \citep{deGennes2004}. For a given system, $\theta$ and $V$ are predefined parameters, $A$ is a constant ($A \in [1, \infty]$) which is uniquely defined by Eq. (\ref{Eq_2_2}). Subsequently, the profile of the liquid-vapor meniscus can be obtained using Eq. (\ref{Eq_2_3}) and Eq. (\ref{Eq_2_4}) (Note: in figure \ref{fig:fig1}, the origin of the coordinate system is not at the center of the solid-liquid contact area, see figure \ref{fig:fig6}). According to Eq. (\ref{Eq_2_3}) and (\ref{Eq_2_4}), we further obtain the width $w$ of the solid-liquid contact area and the height $h$ of the drop,
% ------------------
\begin{eqnarray}
d = \sqrt{2}a \int_0^{\theta} \frac{\cos{\xi}}{\sqrt{A-\cos{\xi}}} {\rm d}{\xi},
\label{Eq_2_5}
\end{eqnarray}
\begin{eqnarray}
h = \sqrt{2}a \left( \sqrt{A - \cos{\theta}} - \sqrt{A-1} \right).
\label{Eq_2_6}
\end{eqnarray}
% ------------------
% ---
\noindent
Moreover, when $\theta \in [0^{\circ}, 90^{\circ}]$, we have $r_{\rm max} = d/2$; when $\theta \in [90^{\circ}, 180^{\circ}]$, we have,
% ------------------
\begin{eqnarray}
r_{\rm max} = \frac{\sqrt{2}a}{2} \int_0^{\pi/2} \frac{\cos{\xi}}{\sqrt{A-\cos{\xi}}} {\rm d}{\xi},
\label{Eq_2_7}
\end{eqnarray}
\begin{eqnarray}
h_1 = \sqrt{2}a \left( \sqrt{A} - \sqrt{A-1} \right), \quad h_2 = \sqrt{2}a \left( \sqrt{A-\cos{\theta}} - \sqrt{A} \right),
\label{Eq_2_8}
\end{eqnarray}
% ------------------
% --
\noindent
where $h_1 = z|_{r=0} - z|_{r=r_{\rm max}}$, $h_2 = z|_{r=r_{\rm max}} - z|_{r=d/2}$ and $h = h_1 + h_2$. A combination of Eq. (\ref{Eq_2_2}) and (\ref{Eq_2_5}) leads to,
% ------------------
\begin{eqnarray}
V = 2a^2 \left[ \frac{\sqrt{2}}{2} \left( \frac{d}{a} \right) \sqrt{A-\cos{\theta}} - \sin{\theta} \right].
\label{Eq_2_9}
\end{eqnarray}
% ------------------

% --
There are two cases which are valuable to be discussed: (1) when $ A \to \infty $, we get $ d \approx \sqrt{2} a \sin \theta / \sqrt{A} $ and $ h \approx \sqrt{2} a (1 - \cos \theta) / 2 \sqrt{A} $ from Eq. (\ref{Eq_2_5}) and (\ref{Eq_2_6}), respectively, which results $ h/d \approx (1-\cos \theta)/2 \sin \theta $. This case corresponds to very small droplets with a spherical shape because the effect of gravity can be ignored; (2) when $A \to 1$, we get $d \to \infty $ and $h \approx 2a \sin(\theta/2)$ \citep{deGennes2004}, which indicates big puddles. In the latter case, when $\theta \in [90^{\circ}, 180^{\circ}]$, Eq. (\ref{Eq_2_8}) reduces to $h_1 \approx \sqrt{2} a$, $h_2 \approx \sqrt{2}a [\sqrt{2} \sin{(\theta/2)} - 1]$. This suggests $h_1$ is approximately constant and $h$ just relies on $h_2$. This essentially implies that if we just focus on the upper part of the liquid (i.e. $z \ge z|_{r=r_{\rm max}}$ in figure \ref{fig:fig1}(b)), we always get a nominal puddle with $\theta = 90^{\circ}$. Moreover, when $A \to 1$ (i.e., $d \to \infty$), the profile of a half puddle (e.g., $x \le 0$) is similar to the meniscus of an infinitely long cylinder pressing at a liquid-air interface, which has received a lot of interest in recent years \citep{Lee2009, Zheng2009}.
% ************
\begin{figure}
  \centerline{\includegraphics[width=13.0cm]{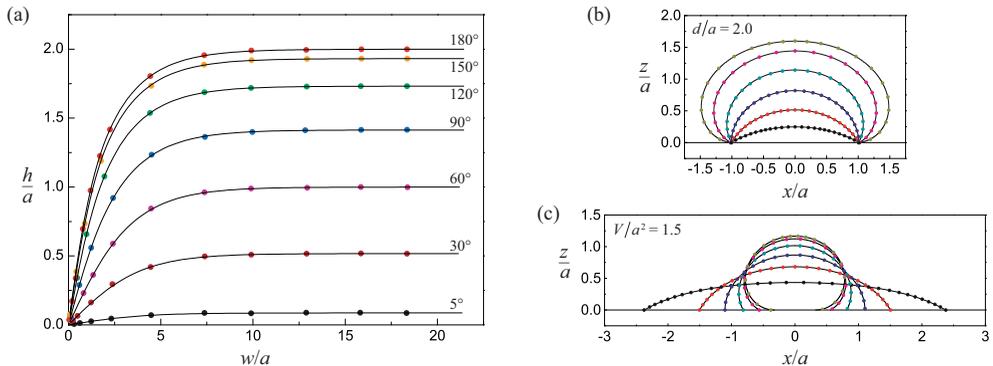}}
  \caption{ Comparisons between theoretical (solid curves) and numerical results (dots): ({\it a}) The dependency of $h/a$ on $w/a$, here $w = 2r_{\rm max}$. The contact angle $\theta$ ranges from $5^{\circ}$ to $180^{\circ}$. When $w/a$ is large enough, $h/a \to 2 \sin (\theta/2)$ \citep{deGennes2004}. ({\it b})({\it c}) Profiles of drops with a fixed solid-liquid contact region $d/a = 2.0$ and a fixed volume $V/a^2 = 1.5$, respectively. Different curves (bottom-up) correspond to $\theta = 30^{\circ}, 60^{\circ}, 90^{\circ}, 120^{\circ}, 150^{\circ}, 180^{\circ}$. The {\it x}-axis and {\it z}-axis shown in ({\it b})({\it c}) are in scale.}
\label{fig:fig2}
\end{figure}
% ************

% --
In order to check the validity of the above theoretical results, we carry out numerical calculations by employing a finite element method ({\it Surface Evolver} \citep{Brakke1992}) and make comparisons between them. In figure \ref{fig:fig2}({\it a}), we give the dependency of $h$ on $w = 2r_{\rm max}$.  Moreover, we also focus on specific cases: we fix the dimensionless values of the solid-liquid contact area at $d/a = 2.0$ and the volume at $V/a^2 = 1.5$ in figure \ref{fig:fig2}({\it b}) and ({\it c}), respectively, but vary the contact angle ($\theta \in [30^{\circ},  180^{\circ}]$). The solid curves represent results obtained using Eq. (\ref{Eq_2_2})-(\ref{Eq_2_4}), and the dots are extracted from {\it Surface Evolver}. These comparisons demonstrate an excellent agreement between each other. 
% --------------------------------------------------------------------------------------------------
%\clearpage
\section{Drops lying on an inclined surface}\label{Sec_3}
\begin{figure}
  \centerline{\includegraphics[width=13.0cm]{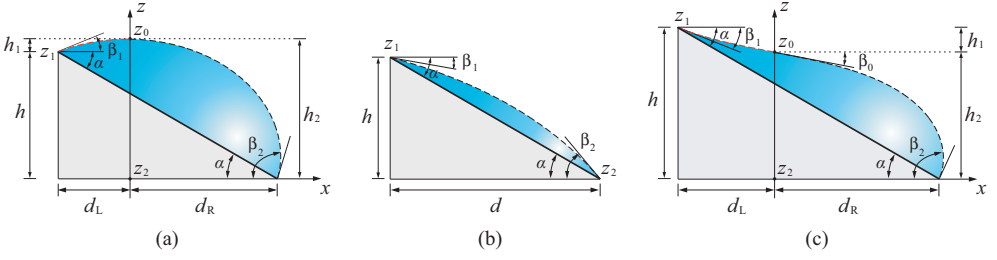}}
  \caption{Definitions of the geometrical parameters for drops lying on inclined surfaces. $\alpha$ is the slope of the surface, $d = d_{\rm L} + d_{\rm R}$ is the projected length of the solid-liquid contact area $d_{\rm SL}$ on the {\it x}-axis and $h = z_1 - z_2$ is the height of $d_{\rm SL}$. We define $\theta_{\rm R} = \beta_1 + \alpha$ and $\theta_{\rm A} = \beta_2 - \alpha$, respectively: ({\it a}) $\theta_{\rm R} \ge \alpha$; ({\it b}) $\theta_{\rm R} \le \alpha$; and ({\it c}) $\theta_{\rm R} \le \alpha$ but the liquid-vapor meniscus consists of a concave and a convex part (more details are given in figure \ref{fig:fig7}).}
\label{fig:fig3}
\end{figure}

By employing the same approach but with modified boundary conditions (Appendix  \ref{appA_3}), we can quantify the wetting state of drops on inclined surfaces. As shown in figure \ref{fig:fig3}, $h$ and $d$ represent the height and the width of the solid-liquid area, respectively. For convenience, we define $\theta_{\rm R} = \beta_1 + \alpha$ and $\theta_{\rm A} = \beta_2 - \alpha$. We have the following three cases. In the first two cases, $\theta_{\rm R} \ge \alpha$ and $\theta_{\rm R} \le \alpha$ (which means $\beta_1 \le 0$) as shown in figure \ref{fig:fig3}({\it a}) and ({\it b}), respectively, the profiles of the liquid-vapor interface are globally convex and can be characterized using the following unified formulas,
% ------------------
\begin{eqnarray}
h = \sqrt{2}a \left( \sqrt{A-\cos{\beta_2}} - \sqrt{A-\cos{\beta_1}} \right),
\label{Eq_3_1}
\end{eqnarray}
% ------------------
\begin{eqnarray}
d = \frac{\sqrt{2}a}{2} \left( \int_{0}^{\beta_1} \frac{\cos{\xi}}{\sqrt{A-\cos{\xi}}} {\rm d}{\xi} + \int_{0}^{\beta_2} \frac{\cos{\xi}}{\sqrt{A-\cos{\xi}}} {\rm d}{\xi} \right).
\label{Eq_3_2}
\end{eqnarray}
% ------------------
% --
\noindent
In fact, we can write $d$ as $d = d_{\rm L} + d_{\rm R}$ with $d_{\rm L} = \frac{\sqrt{2}a}{2} \int_{0}^{\beta_1} \frac{\cos{\xi}}{\sqrt{A-\cos{\xi}}} {\rm d}{\xi}$ and $d_{\rm R} = \frac{\sqrt{2}a}{2} \int_{0}^{\beta_2} \frac{\cos{\xi}}{\sqrt{A-\cos{\xi}}} {\rm d}{\xi}$, and $h$ as $h = h_2 - h_1$  with $h_1 = \sqrt{2}a \left( \sqrt{A-\cos{\beta_1}} - \sqrt{A-1} \right)$ and $h_2 = \sqrt{2}a \left( \sqrt{A-\cos{\beta_2}} - \sqrt{A-1} \right)$. For the case in figure \ref{fig:fig3}({\it b}), $d_{\rm L}$, $d_{\rm R}$, $h_1$ and $h_2$ are virtual geometrical parameters and not shown. In these first two cases, $A \in [1, \infty]$.

% --
However, if the solid-liquid contact area $d_{\rm SL}$ is large enough, as shown in figure \ref{fig:fig3}({\it c}), $\theta_{\rm R} \le \alpha$ (which also means $\beta_1 \le 0$), the liquid-vapor meniscus consists of a concave (on the left) and a convex (on the right) parts. In this case, we obtain, 
% ------------------
\begin{eqnarray}
h = \sqrt{2}a \left( \sqrt{A-\cos{\beta_1}} + \sqrt{A-\cos{\beta_2}}  \right),
\label{Eq_3_3}
\end{eqnarray}
% ------------------
\begin{eqnarray}
d = \frac{\sqrt{2}a}{2} \left( \int_{\beta_0}^{-\beta_1} \frac{\cos{\xi}}{\sqrt{A-\cos{\xi}}} {\rm d}{\xi} + \int_{\beta_0}^{\beta_2} \frac{\cos{\xi}}{\sqrt{A-\cos{\xi}}} {\rm d}{\xi} \right).
\label{Eq_3_4}
\end{eqnarray}
% ------------------
in which $\beta_0 (\ge 0)$ means the slope of the meniscus at $z_0$ (the curvature $\kappa|_{z_0}=0$), and in this case $A = \cos{\beta_0} \in [0,1]$. We can write $d$ as $d = d_{\rm L} + d_{\rm R}$ with $d_{\rm L} = \frac{\sqrt{2}a}{2} \int_{\beta_0}^{-\beta_1} \frac{\cos{\xi}}{\sqrt{A-\cos{\xi}}} {\rm d}{\xi}$ and $d_{\rm R} = \frac{\sqrt{2}a}{2} \int_{\beta_0}^{\beta_2} \frac{\cos{\xi}}{\sqrt{A-\cos{\xi}}} {\rm d}{\xi}$, and $h = h_1 +  h_2$ with $h_1 = \sqrt{2}a \left( \sqrt{A-\cos{\beta_1}} \right)$ and $h_2 = \sqrt{2}a \left( \sqrt{A-\cos{\beta_2}} \right)$. Interestingly, if $d$ is larger than a critical value, an instability should occur and the rear part of the liquid will break into satellite drops \citep{Podgorski2001}, but which is beyond the scope of this paper.

% ************
% ************

% --
Moreover, the volume of the drops in figure \ref{fig:fig3} can be obtained (see Appendix  \ref{appA_3}),
% ------------------
\begin{eqnarray}
\frac{V}{a^2} = \frac{1}{\tan{\alpha}} \left[ \cos{\beta_1} - \cos{\beta_2} \right] - \left[ \sin{\beta_1} - \sin{\beta_2} \right].
\label{Eq_3_5}
\end{eqnarray}
% ------------------
\noindent
Multiplying $\rho g a^2 \sin{\alpha}$ on both sides of Eq. (\ref{Eq_3_5}) leads to Eq. (\ref{Eq_1_1}), which means that Eq. (\ref{Eq_1_1}) is indeed a natural boundary condition of the Young-Laplace equation. 

% --
Next, there are two situations which will be discussed: a completely pinning of the contact line and a movable contact line.
% --------------------------------------------------------------------------------------------------
\subsection{Complete pinning of the contact line}\label{Sec_3_1}
% --
As a consequence of the inevitable roughness of real surfaces, the contact line pinning is a very common phenomenon \citep{deGennes2004}. In this case, for a specific system (with certain values of $V$ and $\alpha$), the solid-liquid contact area $d_{\rm SL}$ is known in advance, so $d = d_{\rm SL} \cos{\alpha}$, $h = d_{\rm SL} \sin{\alpha}$ are also known.  Combine Eqs. (\ref{Eq_1_1}), (\ref{Eq_3_1}) and (\ref{Eq_3_2}) (or Eqs. (\ref{Eq_3_3}) and (\ref{Eq_3_4})) together (recall we have defined $\beta_1 = \theta_{\rm R} - \alpha$ and $\beta_2 = \theta_{\rm A} + \alpha$), the three unknown parameters ($A$, $\theta_{\rm R}$ and $\theta_{\rm A}$) can be found by solving these three equations. 

\begin{figure}
  \centerline{\includegraphics[width=12.0cm]{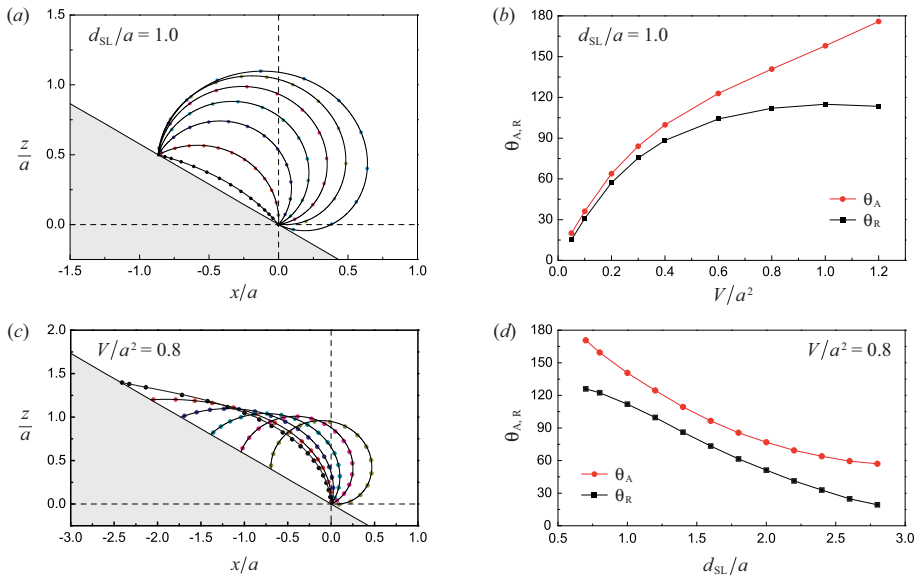}}
  \caption{Profile of drops on inclined surfaces with $\alpha = 30^{\circ}$: ({\it a}) $d_{\rm SL}/a = 1.0$, the volume varies as $V/a^2 = 0.05, 0.2, 0.4, 0.6, 0.8, 1.0, 1.2$; ({\it c}) $V/a^2 = 0.8$, the solid-liquid area varies as $d_{\rm SL}/a = 0.8, 1.2, 1.6, 2.0, 2.4, 2.8$. The solid curves are theoretical results and the dots are numerical results extracted from {\it Surface Evolver}. In ({\it b}) and ({\it d}), the variation of $\theta_{\rm R}$ and $\theta_{\rm A}$ are given, corresponding to ({\it a}) and ({\it c}), respectively. The red dots and black squares are theoretical results with the solid curves as a guide of the eye.}
\label{fig:fig4}
\end{figure}

% --
Two examples are demonstrated in figure \ref{fig:fig4}: ({\it a}), the solid-liquid contact area is fixed at $d_{\rm SL}/a = 1.0$. Different curves correspond to drops with different volumes, i.e. $V/a^2 \in [0.05, 1.2]$; and in ({\it c}), the volume of the drop is fixed at $V/a^2 = 0.8$ with a variation of the solid-liquid contact area $d_{\rm SL}/a \in [0.8, 2.8]$. The solid curves are theoretical results, which agree well with numerical results (dots) extracted from {\it Surface Evolver}. We give the variation of $\theta_{{\rm R}}$ and $\theta_{{\rm A}}$ in figure \ref{fig:fig4}({\it b}) and ({\it d}), corresponding to figure \ref{fig:fig4}({\it a}) and ({\it c}). 

% ************
% ************

The availability of an exact solution of the Young-Laplace equation allows direct evaluation of a range of physical quantities that play an important role in a drop's wetting behaviour. For example, one can calculate the free energy of the drop, which includes two parts, the surface energy $E_{\rm s}$ and gravitational potential $E_{\rm g}$, so $E = E_{\rm s} + E_{\rm g}$. $E_{\rm s}$ is defined using $E_{\rm s} = \sigma (s_{\rm LV} - d_{\rm SL} \cos{\theta})$, in which $s_{\rm LV}$ is the arc length of the liquid-vapor interface. Considering $E_{\rm g}$ depends on relative position, we need a reference level at which to set the potential energy equal to 0. For convenience, we always set the front point of the solid-liquid area at $x = 0, z = 0$, as shown in figure \ref{fig:fig4}({\it a})({\it c}). The reference level will not alter the physics in this problem. Finally, we obtain the normalized total free energy,
% ------------------
\begin{eqnarray}
\nonumber
\frac{E}{a \sigma} & = & \frac{\sqrt{2}}{2} \left( \int_{0}^{\beta_1} \frac{A \cos \xi + \sin^2 \xi}{\sqrt{A-\cos \xi}} {\rm d} \xi + \int_{0}^{\beta_2} \frac{A \cos \xi + \sin^2 \xi}{\sqrt{A-\cos \xi}} {\rm d} \xi \right) + \frac{\sqrt{2}V}{a^2} \sqrt{A - \cos \beta_2} \\
& - & \frac{1}{6} \left( \frac{d}{a} \right)^3 \tan^2 \alpha - \left( \frac{d}{a} \right) \left[ \sqrt{(A - \cos \beta_1)(A - \cos \beta_2)} + \frac{\cos \theta}{\cos \alpha} \right]
\label{Eq_3_6}
\end{eqnarray}
% ------------------
\noindent
for figure \ref{fig:fig3}({\it a}) and ({\it b}). For figure \ref{fig:fig3}({\it c}), we obtain 
% ------------------
\begin{eqnarray}
\nonumber
\frac{E}{a \sigma} & = & \frac{\sqrt{2}}{2} \left( \int_{\beta_0}^{-\beta_1} \frac{A \cos \xi + \sin^2 \xi}{\sqrt{A-\cos \xi}} {\rm d} \xi + \int_{\beta_0}^{\beta_2} \frac{A \cos \xi + \sin^2 \xi}{\sqrt{A-\cos \xi}} {\rm d} \xi \right) + \frac{\sqrt{2}V}{a^2} \sqrt{A - \cos \beta_2} \\
& - & \frac{1}{6} \left( \frac{d}{a} \right)^3 \tan^2 \alpha + \left( \frac{d}{a} \right) \left[ \sqrt{(A - \cos \beta_1)(A - \cos \beta_2)} - \frac{\cos \theta}{\cos \alpha} \right].
\label{Eq_3_7}
\end{eqnarray}
% ------------------
\noindent

% --
From the definition of $E_{\rm s}$ (or Eq. (\ref{Eq_3_6}), (\ref{Eq_3_7})), we know that for a complete contact line pinning case, $d_{\rm SL} \cos \theta$ is a constant, so the Young contact angle has no contribution to determine the profile of the liquid-vapor meniscus. 
% --------------------------------------------------------------------------------------------------
%\clearpage
\subsection{Movable contact line}\label{Sec_3_2}
In this section, we discuss a situation when the drop has a movable solid-liquid contact line. Physically, on the one hand, the drop can adjust its shape and finally reach ``a most likely" wetting state; on the other hand we have to emphasize that ``a movable contact line" indicates that line pinning still exists (but not in a total or partial pinning state), otherwise the drop will continue to slide along the slope due to gravity.
% ************
\begin{figure}
  \centerline{\includegraphics[width=7.0cm]{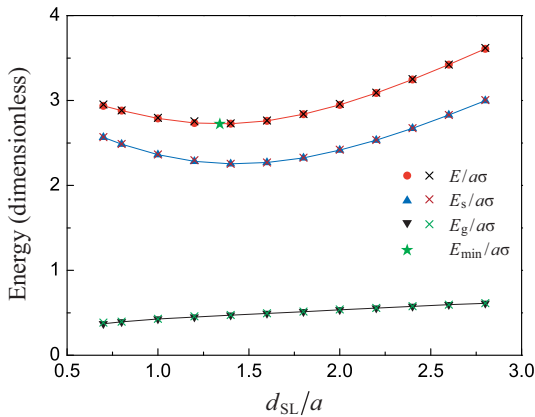}}
  \caption{Dependency of the normalized energy on $d_{\rm SL}/a$. These results correspond to figure \ref{fig:fig4}({\it c}) and ({\it d}), in which $V/a^2 = 0.8$ and $\alpha = 30^{\circ}$. The red circular dots, blue up-pointing and black down-pointing triangles represent numerical results ({\it Surface Evolver}) of $E$, $E_{\rm s}$ and $E_{\rm g}$, respectively. The cross symbols are theoretical results, and the green star indicates $E_{\rm min}$.}
\label{fig:fig5}
\end{figure}
% ************

% --
To determine the most likely wetting state of a specific system ($V$, $\alpha$ and $\theta$ are predefined parameters), we depict figure \ref{fig:fig4} ({\it c}) as an example and give the dependency of $E$ on $d_{\rm SL}$ in figure \ref{fig:fig5}. We can see there is a minimum value (i.e. $E_{\rm min}$) exists, and this state can be exactly characterized by a combination of (based on Eq. (\ref{Eq_3_6}) or (\ref{Eq_3_7}))
% ------------------
\begin{eqnarray}
\frac{{\rm d} E}{{\rm d} d_{\rm SL}} = 0,
\label{Eq_3_8}
\end{eqnarray}
% ------------------
\noindent
and Eqs. (\ref{Eq_1_1}), (\ref{Eq_3_1}) and (\ref{Eq_3_2}) (or Eqs. (\ref{Eq_3_3}) and (\ref{Eq_3_4})). The four unknown parameters (i.e. $A$, $\theta_{\rm R}$, $\theta_{\rm A}$ and $d_{\rm SL}$ (or $d$)) can be thereby uniquely determined by these four equations. We define the correspondingly state as ``the most likely" wetting state. 

% --
Unfortunately, the problem is further complicated by the fact that $A$, $\theta_{\rm R}$, $\theta_{\rm A}$ and $d_{\rm SL}$ are functions of each other and they are coupled, so far we could not express Eq. (\ref{Eq_3_8}) using an explicit matter, we leave this open question for further research. Instead, by employing a numerical way, we can solve these four equations and find the wetting state, we mark the resulting $E_{\rm min}$ and $d_{\rm SL}$ in figure \ref{fig:fig5} using a green asterisk.
% --------------------------------------------------------------------------------------------------
\section{Concluding remarks}\label{Sec_4}
In this letter, we have derived exact analytical solutions of the Young-Laplace equation for 2D drops under gravity, which for the first time is allowing the shape of the drops and other related geometrical parameters (e.g., $h$, $w$, $\theta_{\rm R}$ and $\theta_{\rm A}$) to be fully determined. The excellent agreement demonstrated makes such solutions good candidates in the description of 2D drops beyond the capabilities of the lubrication approximation or other types of perturbation solutions (in powers of $\mathrm {Bo}$ as the small parameter). Although 2D drops are of theoretical (rather than practical) interest, the existence of an exact analytical solution is a potentially useful step for future studies of industrial processes in a 2D case \citep{Schiaffino1997, Gau1999, Xia2012, Reyssat2015}.

% --
We believe that the results presented in this work provide a rather important platform for extensions of a number of fundamental directions in wetting: (1) instead of constant values of $\alpha$ and $V$, we could investigate the dependency of $\theta_{\rm R}$, $\theta_{\rm A}$ and $d_{\rm SL}$ on $\alpha$ or $V$. We believe there are some critical parameters that account for a series of interesting phenomena, such as when the rear contact regime will break into satellite droplets, when the drop will run down the slope, etc.; (2) introducing contact angle hysteresis $\Delta \theta$ and assuming $\theta_{\rm R} = \theta - \Delta \theta/2$ and $\theta_{\rm A} = \theta + \Delta \theta/2$ may give us new perspectives from a different view; (3) since elliptic integrals are widely utilized, we suggest to find emplicit expressions using an asymptotic way built on the exact solutions we have constructed, which would be easier to use and more robust than previous methods which rely on  various approximations (e.g. $z^{\prime} \approx 0$, or $ \mathrm{Bo} \ll 1$).
% --
% --------------------------------------------------------------------------------------------------
\section*{Acknowledgements}
The support of the Alexander von Humboldt Foundation is gratefully acknowledged. 
% --------------------------------------------------------------------------------------------------
\appendix
% --------------------------------------------------------------------------------------------------
\section{Modeling and deduction of the general solution}\label{appA}
% --
Different from the work of \citet[pp. 243]{Landau1987}, in which they only considered a hydrophilic case and the contact angles between the liquid and each side of the two walls are equal (figure \ref{fig:fig6}{\it a}), we extend the discussion to arbitrary contact angles (e.g. $\alpha \in [0^{\circ}, 180^{\circ}]$ and $\alpha_1 \ne \alpha_2 \ne \alpha_3 \ne \alpha_4$). Built on these, we can find exact solutions of the Young-Laplace equation for 2D drops on horizontal and inclined surfaces.
% --------------------------------------------------------------------------------------------------
\subsection{Hydrophilic state}\label{appA_1}
\begin{figure}
  \centerline{\includegraphics[width=9.0cm]{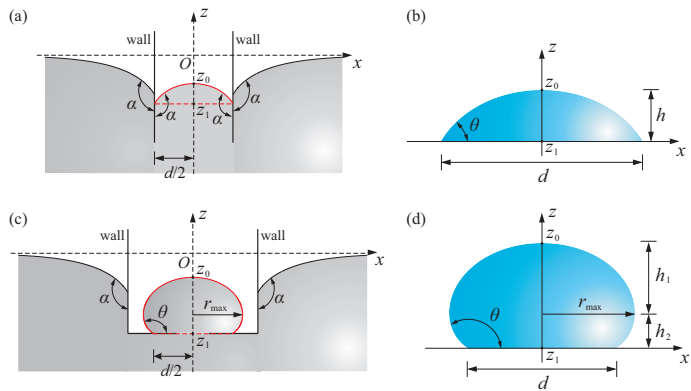}}
  \caption{Wetting and modeling: (a) a 2D meniscus between two plates under gravity. $\alpha = 150^{\circ}$, $d/2 = a$; (b) a 2D drop under gravity. $ \theta = \alpha - \pi/2 = 60^{\circ} $, $d = 2a$; (c) a 2D meniscus between two plates under gravity. $\alpha = 150^{\circ}$ and there is a gap between the bottom walls, a drop is formed with $\theta=\alpha=150^{\circ}$ and $d/2=a$; (d) a 2D drop under gravity. $\theta = 150^{\circ}$, $d = 2a$.}
\label{fig:fig6}
\end{figure}

% --
The key idea is that when we make a comparison between figure \ref{fig:fig6}({\it a}) and \ref{fig:fig6}({\it b}), we can conclude that the shape enclosed by the meniscus between the two walls and the horizontal dashed line in figure \ref{fig:fig6}({\it a}) (as shown in red) is the same as the shape of the 2D drop in figure \ref{fig:fig6}({\it b}) in the case: (1) $ \theta = \alpha - \pi /2 $; (2) the distance between the two walls is equal to the width of the 2D drop. This analysis suggests if we can obtain the profile of the meniscus in figure \ref{fig:fig6}({\it a}), we can get the profile of the 2D drop in figure \ref{fig:fig6}({\it b}).
% ************
% ************

% --
On the basis of the Young-Laplace equation (i.e., Eq. (\ref{Eq_2_1})) and the boundary conditions as shown in figure \ref{fig:fig6}({\it a}) $z|_{x \to \infty} = 0, z^{\prime}|_{x \to \infty} = 0, z^{\prime \prime}|_{x \to \infty} =0 $, we get, 
% ------------------
\begin{eqnarray}
\frac{z}{a^2} - \frac{z^{\prime \prime}}{\left[ 1+ \left( z^{\prime} \right)^2 \right]^{3/2}} = 0.
\label{Eq_A1}
\end{eqnarray}
% ------------------
%--
\noindent
A first integral of Eq. (\ref{Eq_A1}) leads to,
% ------------------
\begin{eqnarray}
\frac{z^2}{2a^2} = A - \frac{1}{\sqrt{1+\left( z^{\prime} \right)^2}},
\label{Eq_A2}
\end{eqnarray}
% ------------------
% --
\noindent
in which $A$ is a constant. We have to emphasize that Eq. (\ref{Eq_A1}) and (\ref{Eq_A2}) are both valid for any part of the meniscus, but here we just focus on the meniscus between the two walls. Regarding $z^{\prime}|_{x=0} = 0$ and $z^{\prime}|_{x=d/2} = 1/\tan{\alpha}$, we can obtain $z_0 = z|_{x=0} = - \sqrt{2}a \sqrt{A-1}$ and $z_1 = z|_{x=d/2} = - \sqrt{2}a \sqrt{A-\sin{\alpha}}$.

% --
By using a transformation \citep[pp. 243]{Landau1987} $ z = - \sqrt{2}a \sqrt{A-\cos{\xi}} $, in which $\xi$ is a variable and $ \xi \in [0, \alpha - \pi/2] $, and replace $\alpha$ by $\theta = \alpha - \pi /2$, we can get the values of $x$, $z$ and $V$ (see Eq. (\ref{Eq_2_2})-(\ref{Eq_2_4})),
% ------------------
\begin{eqnarray}
x = \int_{0}^{\eta} \frac{{\rm d}x}{{\rm d}z} \frac{{\rm d}z}{{\rm d} \xi} {\rm d} \xi = \frac{\sqrt{2}a}{2} \int_{0}^{\eta} \frac{\cos \xi}{\sqrt{A - \cos \xi}} {\rm d}{\xi}, \quad \eta \in [0, \theta],
\label{Eq_A3}
\end{eqnarray}
% ------------------
\begin{eqnarray}
V = 2\int_{0}^{d/2} z {\rm d}x = 2\int_{0}^{\theta} z \frac{{\rm d}x}{{\rm d}z} \frac{{\rm d}z}{{\rm d} \xi} {\rm d} \xi = 2a^2 \left[ \sqrt{A - \cos{\theta}} \int_0^{\theta} \frac{\cos{\xi}}{\sqrt{A - \cos{\xi}}} {\rm d}{\xi}  - \sin{\theta} \right].
\label{Eq_A4}
\end{eqnarray}
% ------------------
% --------------------------------------------------------------------------------------------------
\subsection{Hydrophobic  state}\label{appA_2}
When $\theta \in [90^{\circ}, 180^{\circ}]$, such idea can also be employed: we assume there is a gap between the two bottom walls (see figure \ref{fig:fig6}({\it c})), because of the pressure difference between the middle and the outside walls, there will be a drop formed and its shape (enclosed using the red color in figure \ref{fig:fig6}({\it c})) will be the same as the drop shown in figure \ref{fig:fig6}({\it d}) in the case they have the same values of $\theta$ and $d$. After performing similar calculations as shown in section \ref{appA_2}, we can also obtain Eqs. (\ref{Eq_2_2})-(\ref{Eq_2_9}).
% ------------------
% --------------------------------------------------------------------------------------------------
\subsection{Drop lying on an inclined surface}\label{appA_3}
% ************
\begin{figure}
  \centerline{\includegraphics[width=8.6cm]{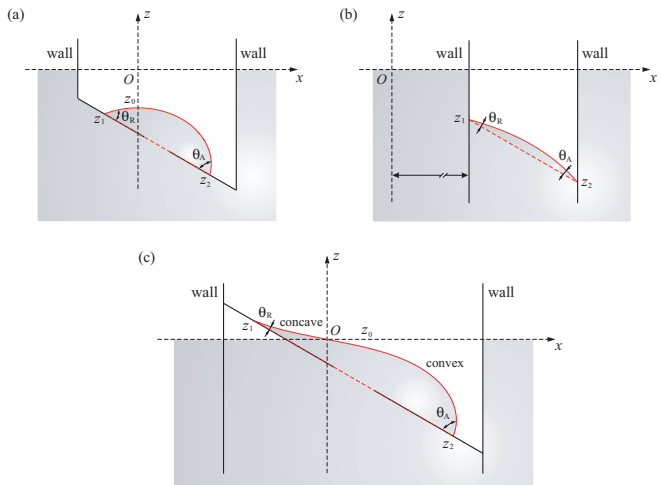}}
  \caption{Modeling and calculations of the menisci. The black solid lines represent solid walls in liquid. There are some gaps between the walls in ({\it a}) and ({\it c}). The red solid and dashed lines represent the liquid-vapor and solid-liquid interface of the virtual drops. $O$ is the origin of the coordinate system. ({\it a}) $\theta_{\rm R} = 51.2^{\circ}$, $\theta_{\rm A} = 76.9$, $d_{\rm SL}/a = 2.0$, $V/a^2 = 0.8$, $z_0 = -0.63$, $z_1 = -0.73$, $z_2 = -1.72$; ({\it b}) $\theta_{\rm R} = 15.3^{\circ}$, $\theta_{\rm A} = 20.0$, $d_{\rm SL}/a = 1.0$, $V/a^2 = 0.05$, $z_1 = -0.40$, $z_2 = -0.90$; ({\it c}) $\theta_{\rm R} = 8.4^{\circ}$, $\theta_{\rm A} = 84.9$, $d_{\rm SL}/a = 4.0$, $V/a^2 = 1.8$, $z_0 = 0$, $z_1 = 0.33$, $z_2 = -1.68$, $\beta_0 = 10.8^{\circ}$. They correspond to ({\it a}),({\it b}),({\it c}) in figure \ref{fig:fig3}, $\alpha = 30^{\circ}$.}
\label{fig:fig7}
\end{figure}
% ************
Lastly, using the similar idea, we model the wetting of drops lying on inclined surfaces, as shown in figure \ref{fig:fig7}. We either use two walls with different contact angles (i.e. figure \ref{fig:fig7}({\it b})) or use an inclined slope between the two walls (with some gap, see figure \ref{fig:fig7}({\it a}), ({\it c})). The virtual 2D drops are enclosed using red curves (the solid and dashed red curves represent the liquid-vapor and solid-liquid interfaces, respectively. $\theta_{\rm R}$ and $\theta_{\rm A}$ are also marked). For convenience, by giving proper contact angles between the liquid and the other parts of the walls, the liquid-vapor menisci are flat, which will not vary the physics.  

% --
The reason for us to employ such modeling is that by this way we can apply the boundary conditions $z|_{x \to \infty} = 0, z^{\prime}|_{x \to \infty} = 0, z^{\prime \prime}|_{x \to \infty} =0$ to Eq. (\ref{Eq_A1}), then we get Eq. (\ref{Eq_A2}) and the other relationships. This is the key difference between our idea and the previous methods for handling this question. On the contrary, if we start modeling directly from a 2D drop, it remains obscure how to proceed.

% --
The other related quantities such as $h$, $d$, $V$ and $E$ of the drops on inclined surfaces as shown in figure \ref{fig:fig3}, \ref{fig:fig4} can also be obtained. For figure \ref{fig:fig3}({\it a}) and ({\it b}), we obtain,
% ------------------
\begin{eqnarray}
\frac{V}{a^2} = \frac{\sqrt 2}{2} \left( \frac{d}{a} \right) \left( \sqrt{A - \cos \beta_1} + \sqrt{A - \cos \beta_2} \right) - \left( \sin \beta_1 + \sin \beta_2 \right),
\label{Eq_A5}
\end{eqnarray}
% ------------------
\begin{eqnarray}
s_{\rm LV} = \frac{\sqrt{2}a}{2} \left( \int_{0}^{\beta_1} \frac{1}{\sqrt{A-\cos \xi}} {\rm d} \xi+ \int_{0}^{\beta_2} \frac{1}{\sqrt{A-\cos \xi}} {\rm d} \xi \right),
\label{Eq_A6}
\end{eqnarray}
% ------------------
\begin{eqnarray}
\nonumber
\frac{E_{\rm p}}{a \sigma} & = & \left( A-\cos \beta_2 \right) \left( \frac{d}{a} \right) - \sqrt{2} \sqrt{A-\cos \beta_2} \left( \sin \beta_1 + \sin \beta_2 \right) - \frac{1}{6} \left( \frac{d}{a} \right)^3 \tan^2 \alpha \\
& + & \frac{\sqrt 2}{2} \left( \int_{0}^{\beta_1} \sqrt{A-\cos \xi} \cos \xi {\rm d} \xi + \int_{0}^{\beta_2} \sqrt{A-\cos \xi} \cos \xi {\rm d} \xi \right).
\label{Eq_A7}
\end{eqnarray}
% ------------------

% --
\noindent
For figure \ref{fig:fig3}({\it c}), we obtain,
% ------------------
\begin{eqnarray}
\frac{V}{a^2} = \frac{\sqrt 2}{2} \left( \frac{d}{a} \right) \left( \sqrt{A - \cos \beta_2} - \sqrt{A - \cos \beta_1} \right) - \left( \sin \beta_1 + \sin \beta_2 \right),
\label{Eq_A8}
\end{eqnarray}
% ------------------
\begin{eqnarray}
s_{\rm LV} = \frac{\sqrt{2}a}{2} \left( \int_{\beta_0}^{-\beta_1} \frac{1}{\sqrt{A-\cos \xi}} {\rm d} \xi+ \int_{\beta_0}^{\beta_2} \frac{1}{\sqrt{A-\cos \xi}} {\rm d} \xi \right),
\label{Eq_A9}
\end{eqnarray}
% ------------------
\begin{eqnarray}
\nonumber
\frac{E_{\rm p}}{a \sigma} & = & \left( A-\cos \beta_2 \right) \left( \frac{d}{a} \right) - \sqrt{2} \sqrt{A-\cos \beta_2} \left( \sin \beta_1 + \sin \beta_2 \right) - \frac{1}{6} \left( \frac{d}{a} \right)^3 \tan^2 \alpha \\
& + & \frac{\sqrt 2}{2} \left( \int_{\beta_0}^{-\beta_1} \sqrt{A-\cos \xi} \cos \xi {\rm d} \xi + \int_{\beta_0}^{\beta_2} \sqrt{A-\cos \xi} \cos \xi {\rm d} \xi \right).
\label{Eq_A10}
\end{eqnarray}
% ------------------
\noindent
Either a combination of Eq. (\ref{Eq_3_1}) and (\ref{Eq_A5}) or Eq. (\ref{Eq_3_3}) and (\ref{Eq_A8}) leads to Eq. (\ref{Eq_3_5}). 
% --------------------------------------------------------------------------------------------------
\bibliographystyle{jfm}
\bibliography{refs}

\end{document}